\documentclass{article}

\input{psfig} 

\title{Computational Geometry Column 35}
\author{%
Joseph O'Rourke\thanks{
Dept. of Computer Science, Smith Col\-lege, North\-ampton, 
MA 01063, USA.
\-orourke@cs.\-smith\-.edu.
Supported by NSF Grant CCR-9731804.
}
}
\date{}
\begin{document}
\bibliographystyle{alpha}
\maketitle
\pagestyle{empty}
\thispagestyle{empty}

\begin{abstract}
The subquadratic algorithm of Kapoor for finding shortest paths
on a polyhedron is described.
\end{abstract}

\begin{figure}[htbp]
\begin{center}
\ \psfig{figure=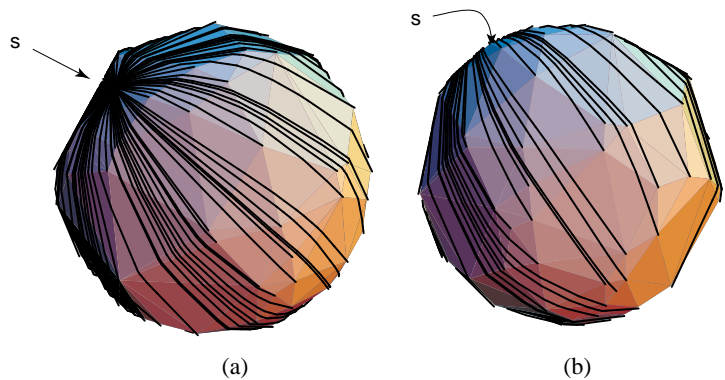,height=5cm}
\end{center}
\caption{Two views of the shortest paths from a
source point $s$ to all $n=100$
vertices of a convex polyhedron~\protect\cite{ox-96};
$s$ is obscured in the ``backside view''~(b).
}
\label{s100}
\end{figure}

A natural shortest paths problem with many applications is:
Given two points $s$ and $t$ on the surface of a polyhedron
of $n$ vertices, find a shortest path on the surface from
$s$ to $t$.  This type of within-surface shortest path is often called
a {\em geodesic shortest path},
in contrast to a {\em Euclidean shortest path}, which may
leave the 2-manifold and fly through 3-space.
Whereas finding a Euclidean shortest path is NP-hard~\cite{cr-nlbtr-87},
the geodesic shortest path may be found in polynomial time.
After an early $O(n^5)$ algorithm~\cite{osb-spps-85},
an $O(n^2 \log n)$ algorithm was developed that
used a technique the authors dubbed the
{\em continuous Dijskstra\/} method~\cite{mmp-dgp-87}.
This simulates the continuous propagation of a {\em wavefront\/}
of points equidistant from $s$ across the surface,
updating the wavefront at discrete events.
It was another decade before this result was improved, by
a clever $O(n^2)$ algorithm that does not track the wavefront
\cite{ch-spp-96}.
This latter algorithm is simple enough to invite implementations,
and several have appeared.
Fig.~\ref{s100} shows an example of using one implementation to
find the shortest paths from $s$ to each vertex
of a convex polyhedron.

Although other geometric shortest path problems saw the breaking
of the quadratic barrier (see~\cite{m-spn-97}),
paths on polyhedra resisted.
One impediment is evident from Fig.~\ref{s100}: 
even on a convex polyhedron, there can
be $\Omega(n^2)$ crossings between polyhedron edges and
paths to the vertices.  So any algorithm that maintains these
paths and treats edge-path crossings as events will be quadratic
in the worst case.
The continuous Dijskstra paradigm faces a similar dilemma:
Examples exist for which there
are $\Omega(n^2)$ wavefront arc-edge crossings.
These obstacles have recently been surmounted by a new
algorithm by Sanjiv Kapoor that 
achieves $O(n \log^2 n)$ time complexity~\cite{k-ecgsp-99}.

Kapoor's algorithm follows the wavefront propagation method,
and is surprisingly similar in overall structure to
the original continuous Dijskstra algorithm~\cite{mmp-dgp-87}.

The algorithm maintains two primary geometric objects
throughout the processing:
the wavefront itself, $W$, which is a sequence of circular arcs,
each centered on either $s$ or a vertex of the polyhedron
(where paths may turn on nonconvex polyhedra);
and a collection $B$ of {\em boundary edges}, edges of the
polyhedron yet to be crossed by the wavefront.
Both of these have size $O(n)$.
Elements of
$W$ and elements of $B$ are related and grouped by
a nearest neighbor relation:
$e \in B$ is associated with arc $a \in W$ if $a$ is closer
to $e$ than to any other arc in $W$.
Boundary edges associated with one arc are grouped into
a {\em boundary section}, and arcs associated with one
boundary edge are grouped into a {\em wavefront section}.
It is this grouping that permits avoiding the quadratic
number of arc-edge crossing events.  The number of
wavefront section-edge events is only $O(n)$.

There remains another quadratic quagmire to be skirted:
Identifying the next event requires computing the distance
from an edge to a wavefront potentially composed of $n$ arcs.
Kapoor handles this by building a hierarchical convex hull
structure for both the wavefront sections and the boundary
sections.  
Subhulls are connected by tangent bridges;
internal nodes store an ``alignment angle'' that 
represents the unfolding relationship between sibling hulls.
These structures permit computing the distance between
a $W$-section and a $B$-section in (essentially)
logarithmic time.
Updating the data structures consumes $O(\log^2 n)$
amortized time per event, which leads to the final
$O(n \log^2 n)$ time complexity.

The details are formidable, and implementation will be
a challenge.  But the many applications and the
significant theoretical improvement suggest implementations
will follow eventually.


\end{document}